# FIRST HIGH POWER PULSED TESTS OF A DRESSED 325 MHZ SUPERCONDUCTING SINGLE SPOKE RESONATOR AT FERMILAB

R. Madrak[#], J. Branlard, B. Chase, C. Darve, P. Joireman, T. Khabiboulline, A. Mukherjee, T. Nicol, E. Peoples-Evans, D. Peterson, Y. Pischalnikov, L. Ristori, W. Schappert, D. Sergatskov, W. Soyars, J. Steimel, I. Terechkine, V. Tupikov, R. Wagner, R. C. Webber, D. Wildman
FNAL, Batavia, IL 60510, U. S. A.

*Abstract*

In the recently commissioned superconducting RF cavity test facility at Fermilab (SCTF), a 325 MHz, $\beta=0.22$ superconducting single-spoke resonator (SSR1) has been tested for the first time with its input power coupler. Previously, this cavity had been tested CW with a low power, high $Q_{ext}$ test coupler; first as a bare cavity in the Fermilab Vertical Test Stand and then fully dressed in the SCTF.

For the tests described here, the design input coupler with $Q_{ext} \sim 10^6$ was used. Pulsed power was provided by a Toshiba E3740A 2.5 MW klystron.

## INTRODUCTION

The Fermilab High Intensity Neutrino Source (HINS) program has been developing 325 MHz superconducting spoke resonators since 2006. Two $\beta=0.22$ (SSR1) cavities have been fabricated. Currently, only SSR1-01, discussed here, has been dressed with a helium vessel and tuners.

The initial HINS design was for a pulsed linac with pulse widths of up to 3 ms and repetition rates of several Hz. Superconducting cavities were to operate at 4 K. More recently, Fermilab has changed strategies for its high intensity linac, Project X, now calling for superconducting cavities operating CW at 2 K. The SSR1 cavities can still be used. The original SSR1 coupler design may also be used with slight modifications.

For future studies, a high power CW 325 MHz source must be procured and the test cryostat must be modified to operate at 2 K. All tests discussed in this paper are at 4.5 K.

## CAVITY AND TEST FACILITY

The initial design of SSR1 cavities for HINS called for 4 K operation at $E_{acc}$ of 10 MV/m with $Q_0 = 5\times10^8$. $E_{acc}$ is defined for the cavity effective length $2/3\beta\lambda = 0.135$ m, the distance between the two outer irises. Details of the design are found in [1]. Results of previous (CW) measurements with the high $Q_{ext}$ coupler are reported in [2]. The maximum $E_{acc}$ for CW operation has been measured to be ~27 MV/m ($Q_0 \sim 3\times10^8$), limited by quench due to field emission. At 10 MV/m $Q_0$ is approximately $10^9$ [2].

In pulsed test mode (these results), the cavity is driven by a 2.5 MW klystron. Klystron output is through WR2300 waveguide, a nominally 10 dB waveguide coupler and 3 inch coaxial waveguide. During cavity testing, power is further limited to $\approx$ 65 kW by attenuators on the klystron RF input. This is sufficient to drive the cavity to its maximum field in approximately 1 ms. Note that since $Q_{ext} \sim 10^6$ and $Q_0 \sim 10^9$, most of the power is reflected.

In the SCTF, the cavity and helium vessel (see Figure 1) are cooled to 4.5 K in a test cryostat [3]. The test cryostat contains a warm magnetic shield and an 80 K thermal shield.

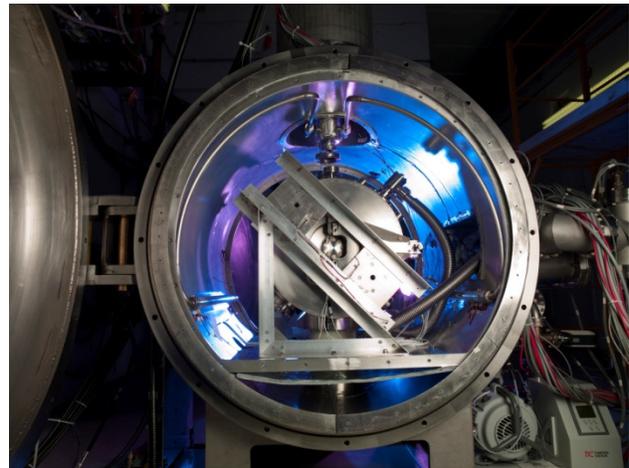

Figure 1: SSR1-01 with helium vessel inside of the 1.44 m OD cryostat. Coils for magnetic field sensitivity studies are mounted on the diagonal frame in front of the vessel.

A new LLRF system was developed to support these tests. It uses an 8-channel receiver to down-convert the cavity probe signal, forward and reflected power signals, and a 325 MHz reference signal to an intermediate frequency of 13 MHz. These four signals are then digitized by a 32-channel multi-field controller (MFC). The MFC uses feed forward, proportional and integral feedback in I and Q to control the cavity amplitude and phase. The controller output is up-converted first to 13 MHZ then to RF frequency by the transmitter. Automatic measurement of the cavity loaded Q and detuning is performed during the decay of the cavity field at the end of the RF pulse.

Pressure fluctuations of ~12 torr and cavity frequency sensitivity of 145 Hz/torr necessitate resonance frequency tracking. The LLRF system uses a 325 MHz reference provided by a signal generator for tunability. A phase locked loop tracks the cavity resonance frequency and adjusts the RF drive frequency accordingly. The PLL has

---

a range of +/-20 kHz. Larger detuning can be handled by first tuning the 325 MHz reference signal. A sample-and-hold approach was implemented to track the cavity resonance during the pulse, yet ignore any detuning happening between pulses. The PLL initial phase value at the beginning of the next pulse is the final value calculated at the end of the previous pulse.

Communication with the MFC is handled by a VXI crate controller running VxWorks. A LabView graphical interface was developed to visualize the controls and readbacks of the cavity field, forward and reflected power.

Fast tuners (piezo actuators in series with slow tuner arms) and associated algorithms for microphonics compensation and Lorentz force detuning compensation are under study as part of this test program [4].

## COUPLER

The SSR1 input coupler is a nominally 50-ohm coaxial design with inner and outer conductor diameters of 33.4 and 78.4 mm. It contains two ceramic windows: one warm and one cold, which protect the integrity of the interior of the cavity and enable the coupler to be divided into warm and cold sections. During cryomodule fabrication, the cold section can be installed on the cavity in the cleanroom prior to assembly of the string. The warm section is installed from outside the vacuum vessel during cryomodule final assembly. The inner conductor is copper and the outer conductor is 304-stainless steel. A short section of hydro-formed bellows in the cold section of the outer conductor allows a small amount of movement – 1 to 2 mm – to be performed from outside the vacuum vessel. This is used to adjust the coupler Q. Figure 2 shows the most recent coupler design. The copper elbow connects to the RF distribution system. The small stainless steel elbow and tube visible in the cross sectional view are the vacuum pumpout for the section between the two ceramic windows. Instrumentation ports are included for vacuum monitoring and multipacting detection.

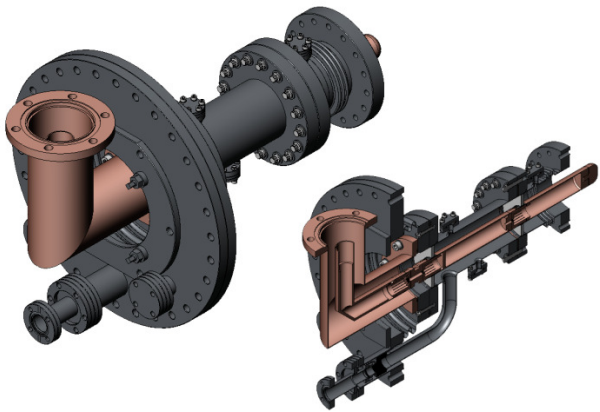

Figure 2: Input coupler assembly and cross sectional view.

### Coupler Tests without Cavity

Before assembly with the cavity, the coupler was tested independently in the HINS 325 MHz RF component test 'cage' using the same klystron that now drives the cavity plus coupler combination. Two couplers were tested at once and were terminated into a 50 ohm water cooled load. The coupler was tested to 500 kW with pulse widths up to 3 ms and a repetition rate of 2 Hz. Multipacting barriers were observed at lower power levels $\sim \leq 20$ kW but these barriers conditioned away.

Once connected to the cavity and installed in the test cryostat, the coupler was also 'warm conditioned' (before cooling down) to 400 kW with the same pulse width and repetition rate. In this case the RF frequency was set to 325.2 MHz which was approximately 450 kHz above the warm cavity resonance frequency (cavity under vacuum and in cryostat insulating vacuum) and outside of the cavity bandwidth.

## CAVITY TESTS

Due to the large overcoupling in this case ($Q_{ext} \sim 10^6$ and $Q_0 \sim 10^9$) $Q_0$ could not be measured. Cavity tests proceeded by gradually increasing the forward power and pulse width, at a 1 Hz repetition rate. Multipacting was observed around 11-13 MV/m. The first quench occurred at 30 MV/m, slightly higher than the maximum $E_{acc}$ of 27 MV/m achieved in CW testing. The field was increased to 34 and then 36 MV/m, during which quenching and large X-ray bursts were observed. After running the cavity for approximately an hour at 36 to 38 MV/m, the X-ray bursts eventually stopped and the quiescent X-ray level also decreased. This strongly suggests that a field emitter had been processed away.

A typical pulse at high flattop field is shown in Figure 3. For faster filling of the cavity, the forward power during the first 1.5 ms of the RF pulse is 1.4 times that during the rest of the pulse. The maximum achieved value of $E_{acc}$ was 40 MV/m with essentially no flattop. Calculations for SSR1 predict that $B_{peak}/E_{acc} = 3.87$ mT/(MV/m) [1]. Thus for these very short pulses the peak surface magnetic field approaches 155 mT.

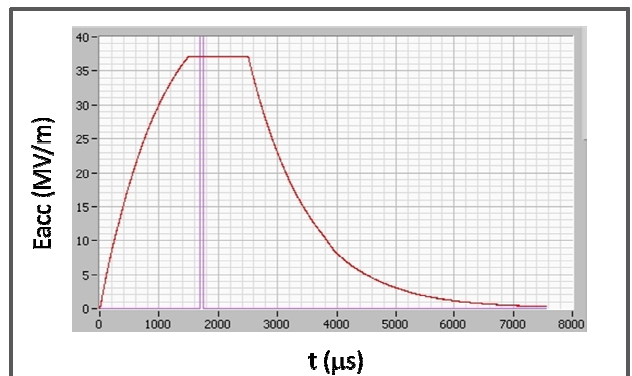

Figure 3: Field vs. time as measured by a field probe and the LLRF system in LabView.

$E_{acc}$ as a function of flattop time was measured. For this study, a step function in forward power was used. The klystron forward power was set to $P_0$ for 1.6 ms; the cavity fill time was $\tau = 2Q_L/\omega_0 \approx 1$ ms. For the remaining time in the RF pulse, forward power was set to $0.73*P_0$. As $P_0$ was increased, maximum field in the cavity and flattop time were measured. In this case, flattop time is defined as the time that the cavity remained on after the initial 1.6 ms without quenching. With $\tau = 1$ ms, the maximum achievable flattop time was limited to $\approx 2$ ms by the klystron power system. Results are displayed in Figure 4.

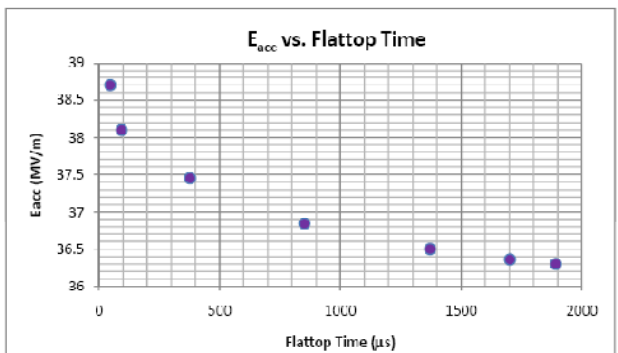

Figure 4: $E_{acc}$ vs. flattop time.

*Sensitivity to Magnetic Fields*

The Project X Linac design calls for solenoidal focusing. The fringe magnetic field requirement for the lenses was established to be at the level of 10 μT. This was subject to some criticism as being both too low and too high. To clarify the issue, several tests were conceived and performed. To study the quality factor degradation due to the presence of magnetic field, the cavity was quenched repeatedly at 37.5 MV/m in the presence of a field, which was generated by four coils mounted inside of the cryostat (see Figure 1). The coils were placed just outside the helium vessel endwalls and were aligned with the spoke. The magnetic field was measured by a Hall probe located inside of the cryostat but outside of the helium vessel. A thick aluminium frame provided a good thermal connection between the coils and the LN2 shield of the cryostat. A platinum resistance thermometer was used to measure the temperature of the coils. Because the first test using a DC excitation current (up to 4 A, where the runaway started) showed no degradation [2], the coils were pulsed to 8 A at 1 Hz with a 200 ms pulse length. A model which takes into account the fact that the field must penetrate the thick stainless steel wall of the cavity helium vessel indicates that the field at the cavity end wall reaches ~20 G during the quench. Figure 5 shows the measured Hall probe signal and the coincident RF pulse seen (due to cross-talk) as a spike.

The quenching at 37.5 MV/m was repeated 25,000 times. Any measureable effect would be seen as a drop in the cavity $Q_0$ as a function of number of quenches. $Q_0$ could not be measured in this coupler configuration, but the effect can also be seen as a drop in quench field. All previous measurements have shown that a drop in quench field coincides with a drop in $Q_0$. No such drop in quench field was observed, indicating that the quenches were not at the endwall. Further studies with high magnetic field at the spokes are being planned.

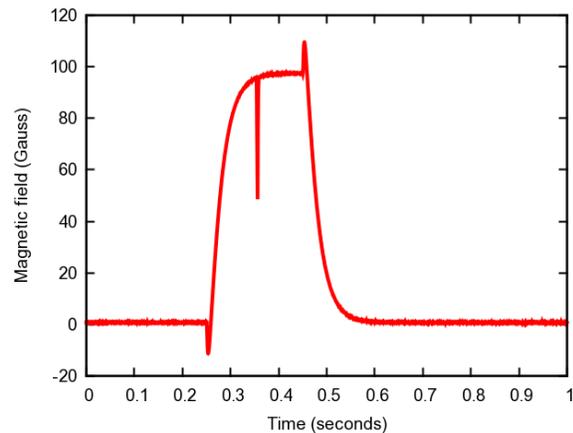

Figure 5: Hall probe signal with spike due to cross-talk with RF pulse.

## FUTURE PLANS

In the near future, SSR1 in SCTF will continue to be used for fast tuner tests and further magnetic field sensitivity studies. Plans are underway to extend SCTF operation to 2K in FY11. Once this is accomplished, studies described in [2] and in this paper can be repeated. SSR1-02 and twelve new SSR1 cavities, expected to arrive at Fermilab in 2011, are ultimately destined for testing at SCTF.

## ACKNOWLEDGEMENTS

We would like to sincerely thank Mark Dilday, Elias Lopez, Kyle Kendziora, Wade Muranyi, Ryan Montiel and Brad Tennis for their patience and excellent work on this project.